\documentclass[12pt]{iopart}

\usepackage{iopams}\usepackage{latexsym,graphicx} 
\begin{document}

\newcommand{\kt}[1]{\ensuremath{|#1\rangle}}
\renewcommand{\br}[1]{\ensuremath {\langle #1|}}
\newcommand{\HS}{\mathcal{H}}
\newcommand{\bk}[2]{\ensuremath {\langle #1|#2 \rangle}}
\newcommand{\ckt}[1]{\ensuremath{|#1\}}}
\newcommand{\cbr}[1]{\ensuremath {\{#1|}}

\title[Entanglement in Massive Coupled Oscillators]{Entanglement in Massive Coupled Oscillators}

\author{N.L.~Harshman and W.F.~Flynn}

\address{Department of Physics,
4400 Massachusetts Ave. NW, American University, Washington, DC 20016-8058}
\ead{harshman@american.edu}
\begin{abstract}
This article investigates entanglement of the motional states of massive coupled oscillators.  The specific realization of an idealized diatomic molecule in one-dimension is considered, but the techniques developed apply to any massive particles with two degrees of freedom and a quadratic Hamiltonian.  We present two methods, one analytic and one approximate, to calculate the interatomic entanglement for Gaussian and non-Gaussian pure states as measured by the purity of the reduced density matrix. The cases of free and trapped molecules and hetero- and homonuclear molecules are treated.  In general, when the trap frequency and the molecular frequency are very different, and when the atomic masses are equal, the atoms are highly-entangled for molecular coherent states and number states.  Surprisingly, while the interatomic entanglement can be quite large even for molecular coherent states, the covariance of atomic position and momentum observables can be entirely explained by a classical model with appropriately chosen statistical uncertainty.
 \end{abstract}

\maketitle

\section{Introduction}

Harmonically-coupled massive oscillators provide a suitable model for many physical systems that are employed or proposed for quantum information processing with continuous variables.  For example, both the longitudinal modes~\cite{james_quantum_1998} and the transverse modes~\cite{zhu_trapped_2006} of ions in Paul traps can be treated (in certain regimes, at least ap\-prox\-i\-mately) as coupled oscillators.  Interactions between oscillators, either direct or mediated by external elements or fields, lead to entanglement between the oscillators that in principle could  be externally controlled or extracted.  In particular, the characterization of entanglement in harmonic chains~\cite{audenaert_entanglement_2002}, one-dimensional arrays of coupled harmonic oscillators, is a paradigm that has attracted sustained attention.  Beside quantum information processing, an additional motivation for investigating systems with harmonic lattice Hamiltonians is the study of the role of entanglement in phase transitions~\cite{amico_entanglement_2008}.

Most research on massive coupled oscillators has employed Gaussian states, the workhorse of continuous variable quantum information theory.  This extensive use of Gaussian states is physically motivated: many dynamical processes result in Gaussian states.  In particular, the ground states and thermal states of massive coupled oscillators with quadratic Hamiltonians are exactly Gaussian.  Their use is also theoretically convenient because many results for Gaussian states in quantum optics can be carried over to massive oscillators.  More generally, there exists a correspondence (up to local coordinate transformations) between covariance matrices and Gaussian states that allows for deep mathematical analysis, including complete characterization of bipartite entanglement and partial characterization of multipartite entanglement~\cite{adesso_entanglement_2007,serafini_standard_2007}.  In contrast, one goal of this paper to to provide methods for studying the entanglement of \emph{non-Gaussian} continuous variable states.  Advances in theory and experiment have led to an increased interest in quantum information processing with non-Gaussian states, and this work investigates the simplest special case of two-mode pure states. 

Our results apply directly to any case of massive particles with two degrees of freedom and a Hamiltonian quadratic in position, but for conceptual ease and clarity we structure our discussion in the language of a diatomic molecule: two distinguishable atoms interacting via a quadratic potential in one-dimension.  This can be thought of as an example of the shortest harmonic chain, but the analysis will not be restricted to the case of equal masses as is usually considered in such systems.  Two different Hamiltonians are considered simultaneously: a molecule trapped in a harmonic potential and an untrapped molecule with a Gaussian wave packet.  The properties of two-mode Gaussian states are well-known from a variety of contexts in quantum optics and continuous-variable quantum information theory~\cite{fan_eigenvectors_1994,simon_peres-horodecki_2000,botero_modewise_2003,marian_gaussian_2008,pirandola_correlation_2009}. For the ground state and coherent states, we will translate our model into the standard language of covariance matrices and logarithmic negativity.  However, for non-Gaussian pure states, the entanglement will be quantified in terms of the purity of the reduced density matrix.  Two methods, one more suitable for analytic calculations and one more suitable for numerical simulations, will be provided.

Entanglement is a notion that depends on the observables one uses to describe a system~\cite{zanardi_quantum_2004}, and the coupled oscillator model allows this connection to be explored in an exactly solvable model.  Because of the coupling interaction, the ``normal'' molecular observables are more convenient for studying and controlling the system properties and dynamics than the the ``natural'' atomic observables.  However, one could imagine that atomic observables like position and momentum are still physically accessible, perhaps through some independent coupling with internal atomic structure.  Assuming that both atomic observables and molecular observables can form a complete set of operationally-accessible interactions and measurements, one can talk about entanglement with respect to the tensor product structure induced by either the molecular set or the atomic set of observables (see also the discussion in Ref.~\cite{torre_entanglement_2010}).  Additionally, by studying the local unitary operators acting on the atomic tensor product structure, one can identify equivalence classes of equally-entangled states and classes of Hamiltonians that lead to entanglement-equivalent dynamics.  For example, these methods demonstrate immediately that coherent states and the ground state have the same entanglement between atoms, and that the dynamics of the entanglement is independent of the linear terms in the Hamiltonian.

We can also interpret our results for the entanglement of atoms in a diatomic molecule as an example of continuous-variable entanglement constrained by conservation laws.  In this perspective, we find that the entanglement between massive oscillators has two determining factors, one dynamic and one kinematic in origin.  First, the Hamiltonian is diagonal in the molecular observables, so the center-of-mass and relative modes are not mixed by the dynamics.  As a result,  the ground state, coherent states, and number states are separable in the molecular observables.  Some dynamic parameter, e.g.~the ratio of the molecular frequency to the trap frequency, will therefore set one scale in the analytic formulas for entanglement.   Additionally, the transformation between atomic and molecular coordinates induces a kind of purely kinematic squeezing of the two-particle wave function.  The motional entanglement created by this kind of wave packet squeezing is mathematically similar to phase space squeezing in quantum optics.  This effect was perhaps first noted by Fan and Klauder~\cite{fan_eigenvectors_1994} who, inspired by the original EPR paper, studied eigenstates of the relative motion of two-particle systems and constructed two mode entangled states in analogy to photonic two mode squeezed states~\cite{hongyi_relationship_2003}.  Subsequent work by Fan generalized these considerations to the entangled state representation for two unequal masses~\cite{fan_hong-yi_common_1995,fan_solvingtwo-body_1996}.
A similar entanglement mechanism occurs in the reflected modes of two-particle scattering systems~\cite{schmuser_entanglement_2006,harshman_entanglement_2008-1} and the same mechanism can be found for wave packet entanglement in photoionization~\cite{fedorov_packet_2004}, spontaneous emission~\cite{fedorov_spontaneous_2005}, and other disassociation processes~\cite{fedorov_short-pulse_2006}.

In terms of diatomic systems, possible physical realizations could include two ions in a linear Paul trap or cold polar diatomic molecules in an optical trap. For example, entanglement swapping between internal atomic degrees of freedom and molecular degrees of freedom has already been demonstrated for two pairs of oscillating ions in a linear Paul trap~\cite{jost_entangled_2009} and novel schemes for entangling transverse modes ion traps have been proposed~\cite{serafini_manipulatingquantum_2009}.  A straightforward implementation scheme for measuring this kind of entanglement would require independent access and measurement of both molecular and atomic canonical observables (i.e., position and/or momentum).  Alternatively, if the molecule could be disassociated with a strong pulse that does not change the original spatial distribution of the wave packet very much, the ratio of the width of the conditional wave packet to the single particle wave packet takes exactly the same value as the purity of the reduced density matrix~\cite{fedorov_packet_2004,fedorov_short-pulse_2006}.  However, either approach would require measurement resolution finer than the scale of the wave packet variation.

Unfortunately, as discussed in the conclusion, even if two-particle spatial covariance measurements were accurate enough to quantify the interatomic entanglement, such measurements cannot establish the `quantumness' of the correlations.  The correlations between atoms revealed by spatial measurements does not exceed those that are possible in some classical system with statistical correlations.  To overcome this, several schemes for developing Bell-type inequalities have been proposed for detecting entanglement in continuous variable systems, such as displaced parity operators~\cite{banaszek_nonlocality_1998} and pseudospin operators~\cite{chen_maximal_2002}.  While these are useful theoretical  discriminators of non-classical correlations, these schemes would appear to require full state tomography to reconstruct the correlations between arbitrary two-mode states~\cite{chen_maximal_2002}.  Disassociation-time entanglement has also been proposed to measure motional entanglement of two atoms disassociated from a diatomic molecule~\cite{gneiting_entanglingfree_2009}, but there the measured entanglement would be created by the disassociation pulse, and not the initial entanglement of the bound state, which is what interests us here.

An additional physical motivation for this work is to study entanglement in bound states of strongly-interacting particles.  For example, because the Moshinsky atom~\cite{moshinsky_good_1968}, a coupled oscillator model for two-electron atoms, is analytically solvable, it has been used to test and explore approximation schemes for multi-electron settings like Hartree-Fock~\cite{moshinsky_good_1968,oneill_wave_2003,march_proposed_2008} and density functional theory~\cite{neal_density_1998,ragot_exact_2006,jens_peder_dahl_moshinsky_2009}.  The amount of spatial entanglement between electrons in the Moshinsky atom (and also in the related Hooke's atom model) has been shown to be a good proxy for the deviations in energy prediction entailed by the separability assumptions used in Hartee-Fock~\cite{amovilli_quantum_2004,march_kinetic_2006,yaez_quantum_2010} and density functional theory~\cite{coe_entanglement_2008,pipek_measures_2009}.  The results presented in this article reproduce the direct calculations of the spatial entanglement of the ground state found in Ref.~\cite{pipek_measures_2009} and of the lowest energy eigenstates in Ref.~\cite{yaez_quantum_2010} (when adjusted to account for the difficulties associated with entanglement of identical fermions).  Our results also explain an entanglement symmetry first noticed in Ref.~\cite{yaez_quantum_2010}: when the center-of-mass and relative energy scales are interchanged, the entanglement is invariant.

The structure of this article is as follows.  First, the model for the diatomic molecule is introduced, and some relevant properties of molecular and atomic entanglement are established.  Then the atomic entanglement of coherent molecular states, including the ground state, is calculated and compared to known results for Gaussian states established using covariance matrix methods.  In the next section, one method for generating the exact expression for the atomic entanglement in an arbitrary energy eigenstate is presented, and an alternate, approximate procedure is described in Appendix A.  The final section discusses the interpretation of the correlations implied by atomic entanglement and how the covariance in this quantum system compares to a classical, statistical description of the same system.

\section{Idealized Diatomic Molecule}

The Hamiltonian for the trapped diatomic molecule can be expressed in terms of the ``normal'' or  molecular center-of-mass observables $\{\hat{X}, \hat{P}\}$ and relative observables $\{\hat{R}, \hat{Q}\}$
\begin{equation}\label{eq:molHam}
\hat{H} = \frac{1}{2M}\hat{P}^2 + \frac{1}{2\mu}\hat{Q}^2 +\frac{1}{2}M\Omega^2\hat{X}^2 + \frac{1}{2}\mu\omega^2\hat{R}^2,
\end{equation}
where $M=m_1 +m_2$ is the total mass, $\mu=m_1 m_2/M$ is the reduced mass.  These observables are related to the ``natural'' atomic observables by the symplectic transformation
\begin{eqnarray}  \label{eq:Observables Definition}
\hat{X} &=& \mu_1\hat{X}_1 + \mu_2\hat{X}_2,\ 
\hat{R} = \hat{X}_1 - \hat{X}_2 - \ell,\nonumber\\
\hat{P} &=& \hat{P}_1 + \hat{P}_2,\
\hat{Q} = \mu_2\hat{P}_1 - \mu_1\hat{P}_2,
\end{eqnarray}
where $\mu_i = m_i/M$ are the mass fractions and $\ell$ is the equilibrium length of the molecule.  The equilibrium length (or the coefficient of any linear term in the Hamiltonian) can be set equal to zero without changing entanglement because such a redefinition corresponds to a local unitary transformation in either molecular or atomic coordinates, as discussed below.  Written in the atomic observables with $\ell=0$, the Hamiltonian becomes
\begin{eqnarray}\label{eq:atoHam}
\hat{H} = \frac{1}{2m_1}\hat{P}_1^2 + \frac{1}{2m_2}\hat{P}_2^2 + \hat{V}(\hat{X}_1,\hat{X}_2),\ \mbox{where}\\
\hat{V} = \frac{1}{2}M\Omega^2\left(\mu_1^2 + \mu_1\mu_2 g^2\hat{X}_1^2 + \mu_2^2 + \mu_1 \mu_2 g^2\hat{X}_2^2+\mu_1\mu_2(1- g^2)\hat{X}_1\hat{X}_2\right).\nonumber
\end{eqnarray}
In the last line we have expressed the ratio of the molecular frequency to the trap frequency as $g=\omega/\Omega\,$ and we will interpret all results for the bound molecule in terms of this dynamical scale parameter.  

Looking at the Hamiltonian in the two different coordinate systems, the benefit of the molecular observables is clear (and well-known).  Cast into the language of entanglement, one can say the Hamiltonian is a separable operator in the molecular Hilbert space partition $\mathcal{H}=\mathcal{H}_{\mathrm{r}}\otimes \mathcal{H}_{\mathrm{c}}$  induced by the relative (denoted ``r'') and center-of-mass observables (``c'')
\begin{eqnarray}
\hat{H} &=& \hat{H}_{\mathrm{r}}\otimes \mathbb{I}_{\mathrm{c}}  + \mathbb{I}_{\mathrm{r}} \otimes \hat{H}_{\mathrm{c}}\nonumber\\
&=& \hbar\omega(\hat{a}\hat{a}^\dag+1/2)\otimes \mathbb{I}_{\mathrm{c}}+ \mathbb{I}_{\mathrm{r}} \otimes\hbar\Omega(\hat{b}\hat{b}^\dag+1/2).
\end{eqnarray}
In the last line the Hamiltonian has been written in terms of  the ladder operators for the molecule oscillations $\{\hat{a},\hat{a}^\dag\}$ and for the trap oscillations  $\{\hat{b},\hat{b}^\dag\}$:
\begin{eqnarray}
\hat{a} &=& \frac{\gamma}{\sqrt{2}}\hat{R} + \frac{i}{\sqrt{2}\hbar\gamma}\hat{Q}\nonumber\\
\hat{b} &=& \frac{\Gamma}{\sqrt{2}}\hat{X} + \frac{i}{\sqrt{2}\hbar\Gamma}\hat{P},
\end{eqnarray}
where $\Gamma = \sqrt{M\Omega/\hbar}$ and $\gamma = \sqrt{\mu\omega/\hbar}=\sqrt{\mu_1\mu_2g}\,\Gamma$ are proportional to the momentum uncertainties of the ground state.  In contrast, the Hamiltonian expressed in atomic observables (\ref{eq:atoHam}) does not separate with respect to the atomic tensor product structure $\mathcal{H}=\mathcal{H}_1\otimes \mathcal{H}_2$ unless $g=1$ (or the unphysical case of infinite mass imbalance, $\mu_1$ or $\mu_2\rightarrow 0$).  When $g=1$, the molecular and trap frequencies are the same and the term proportional to $\hat{X}_1\hat{X}_2$ vanishes (although $\gamma$ and $\Gamma$, which depend on the masses, may still be different).

The energy eigenstates $\kt{m,n}=\kt{m}\otimes\kt{n}$ have the standard harmonic oscillator wave functions $\Phi_{m,n}(r,x)=\phi_m(r)\phi_n(x)=\bk{r,x}{m,n}$ when expanded on the spectrum of $\{\hat{R}, \hat{X}\}$:
\begin{eqnarray}\label{form}
\phi_m(r)&=&\left(\frac{\gamma^2}{\pi}\right)^{1/4} (2^{m} m!)^{-1/2} H_m(\gamma r) e^{-\gamma^2r^2/2}\nonumber\\
\phi_n(x)&=&\left(\frac{\Gamma^2}{\pi}\right)^{1/4} (2^{n} n!)^{-1/2} H_n(\Gamma x) e^{-\Gamma^2x^2/2}
\end{eqnarray}
The energy eigenstates are separable, and therefore unentangled, with respect to the molecular tensor product structure $\mathcal{H}_{\mathrm{r}}\otimes \mathcal{H}_{\mathrm{c}}$, although out of these basis vectors one can construct combinations that are entangled in the molecular tensor product structure, e.g.\ $\kt{m,n}\pm\kt{n,m}$.  One can also define two-mode molecular coherent states
\begin{eqnarray}\label{molcoh}
\kt{\alpha, \beta} & \equiv & D(a,\alpha)D(b,\beta)\kt{0,0}\nonumber\\
&=& e^{-|\alpha|^2/2 -|\beta|^2/2}\sum_{m,n=0}^{\infty}\frac{\alpha^m\beta^n}{\sqrt{m! n!}} \kt{m,n}
\end{eqnarray}
which have well-known physical interpretations as the ``most classical'' harmonic oscillator states (see, for example \cite{schleich_quantum_2001}) and will be useful for subsequent calculations.  The complex number $\alpha$ is the displacement of the relative $\hat{a}$ mode and $\beta$ is the displacement of the center-of-mass $\hat{b}$ mode.

Any state $\kt{\Phi}$ can also be represented by wave functions on the spectrum of the atomic position observables:
\begin{equation}
\tilde{\Phi}(x_1,x_2) =\{x_1,x_2|\Phi\rangle= \Phi(x_1-x_2,\mu_1 x_1 + \mu_2 x_2)
\end{equation}
where we have used the ``curly ket'' notation $|x_1,x_2\}$ to indicate that these are generalized eigenvectors of the atomic position observables, as opposed to the molecular position observable eigenkets $\kt{r,x}$. For almost all values of kinematic and dynamic parameters, and for almost all quantum numbers and superpositions, the wave function is not separable, i.e.~$\tilde{\Phi}(x_1,x_2)\neq\tilde{\phi}_1(x_1)\tilde{\phi}_2(x_2)$ for any $\tilde{\phi}_i$.  This entanglement is evident from the contour plots of probabilities densities in $x_1,x_2$-space depicted in Figures 1 and 2.  Only the top two contour plots of Figure 1 depict separable states, which can be recognized because all marginal probabilities for a given value of one coordinate (say, $x_1$) take the same functional form in the other coordinate.  Or more qualitatively, the ``principle axes'' of the probability densities line up with the coordinate axes when the function is separable.

\begin{figure}
\centering
\includegraphics{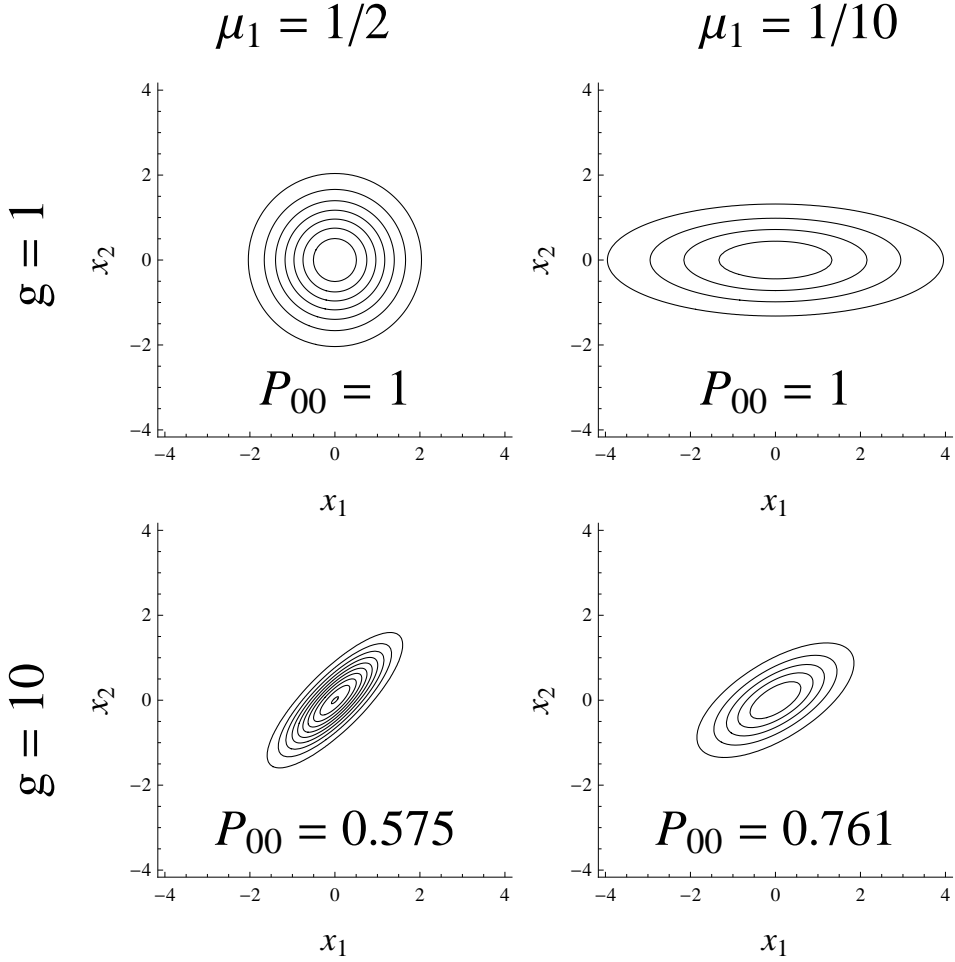}
\caption{These contour plots depict the atomic position probability densities $|\tilde{\Phi}_{00}(x_1,x_2)|^2$ for four combinations of values for $g$ and $\mu_1$.  Positions are measured in units of $\Gamma$.  Note that $\mu_2 = 1-\mu_1.$}
\label{fig1} 
\end{figure}

\begin{figure}
\centering
\includegraphics{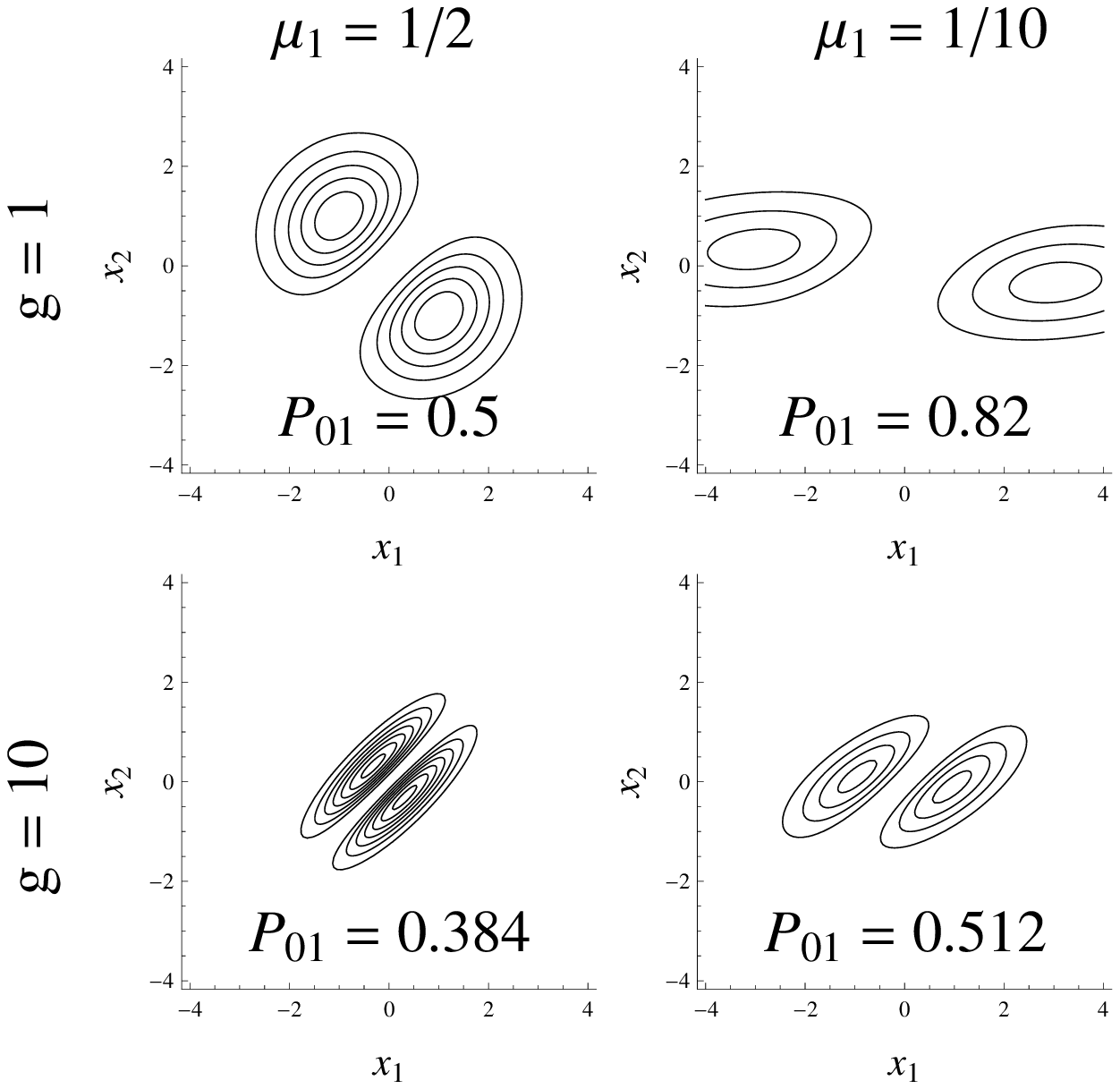}
\caption{These contour plots depict the atomic position probability densities $|\tilde{\Phi}_{10}(x_1,x_2)|^2$ for four combinations of values for $g$ and $\mu_1$. Positions are measured in units of $\Gamma$. Note that $\mu_2 = 1-\mu_1.$}
\label{fig2} 
\end{figure}

To calculate the entanglement between atoms for arbitrary pure states, we will use the purity of the reduced matrix element $P$ (or purity, for short):
\begin{eqnarray}\label{eq:pur}
P=\Tr_1\hat{\rho}_1^2 &=& \int dx_1 \cbr{x_1}\hat{\rho}_1^2\ckt{x_1}\nonumber\\
&=& \int dx_1 dx'_1 \cbr{x_1}\hat{\rho}_1\ckt{x'_1}\cbr{x'_1}\hat{\rho}_1\ckt{x_1}\nonumber\\
&=& \int dx_1 dx'_1dx_2 dx'_2\tilde{\Phi}(x_1,x_2)\tilde{\Phi}^*(x'_1,x_2)\tilde{\Phi}(x'_1,x'_2)\tilde{\Phi}^*(x_1,x'_2)
\end{eqnarray}
where
\begin{eqnarray}
\hat{\rho}_1 &=& {\rm Tr}_2(\kt{\Phi}\br{\Phi})\nonumber\\
& =& \int dx_1 dx'_1dx_2 \tilde{\Phi}(x_1,x_2)\tilde{\Phi}^*(x_1,x_2)\ckt{x_1}\cbr{x'_1}.
\end{eqnarray}
The purity of the reduced density matrix is an entanglement measure for pure states and lies in the range $(0,1]$, with lower values meaning more entanglement (some prefer to use the linear entropy $L=1-P$ for this reason).  For comparison, in $d$-level discrete systems the purity is bounded from below by $d^{-1}$.

As with all reasonable entanglement measures on pure states, the value of $P$ should not change when the state is transformed by a unitary transformation that is separable with respect to tensor product structure.  In particular, operators that are exponentials of linear combinations of atomic or molecular position and momentum observables, such as
\begin{equation}
e^{(i/\hbar(a \hat{R} + b \hat{X}+ c \hat{Q}+ d \hat{P}))},\ e^{(i/\hbar(a \hat{X}_1 + b \hat{X}_2+ c \hat{P}_1+ d \hat{P}_2))},\ \mbox{or}\ D(a,\alpha)D(b,\beta)
\end{equation}
 are separable with respect to both the atomic tensor product structure $\mathcal{H}_1\otimes \mathcal{H}_2$ and molecular tensor product structure $\mathcal{H}_{\mathrm{r}}\otimes \mathcal{H}_{\mathrm{c}}$.  Therefore translations of either the molecular or atomic coordinate systems do not affect the amount of entanglement.  This explains why the equilibrium length of the molecule can be set to $\ell=0$ with out changing any conclusions about entanglement.  The time evolution operator on energy eigenstates is just a phase, so one can also see that stationary states have constant entanglement in time according to any tensor product structure.  Finally, Fourier transforms are also local unitary operators.  Therefore, one can work in atomic momentum space or atomic position space and calculate the same value for the entanglement between the atoms.  We choose to work in position space.

In the limit $\Omega\rightarrow 0$, the term in the Hamiltonian (\ref{eq:molHam}) that leads to the center-of-mass trapping vanishes and our system becomes an untrapped diatomic molecule.  We can still consider wave functions (\ref{form}) of the form $\Phi_{m0}(r,x)=\phi_m(r)\phi_0(x)$, but these are no longer energy eigenstates. The constant $\Gamma$ now plays the role of an initial condition, not a dynamical parameter as in the trapped case.  The quantity $\hbar\Gamma/\sqrt{2}$ can now be interpreted as the momentum uncertainty of the center-of-mass Gaussian wave packet at the moment in time (say $t=0$) when  the wave packet satisfies the minimum uncertainty relation $\Delta x \Delta p=\hbar/2$.  Of course, the center-of-mass wave packet will spread as a function of time
\begin{equation}
\phi_u(x,t)= \frac{1}{\sqrt{1 + i \tau}}\left(\frac{\Gamma^2}{\pi}\right)^{1/4} \exp\left[\frac{-\Gamma^2 x^2}{2(1+\tau^2)}(1 + i\tau)\right]
\end{equation}
where we use a unitless rescaled time $\tau = \Gamma^2\hbar t/M$.  Translations in center-of-mass position and momentum (or equivalently, changes in reference frame) do not affect the entanglement for the unbound molecule, as can be seen from the preceding argument.  Therefore, the entanglement of the $m$th molecular vibrational state with a Gaussian center-of-mass momentum distribution can be calculated from $\tilde{\Phi}(t)_{mu}(x_1,x_2)=\phi_m(x_1-x_2)\phi_u(\mu_1 x_1 + \mu_2 x_2,t)$.  Note that since these states are no longer energy eigenstates, atomic entanglement will not be constant.  In the next section we show that the entanglement  for the unbound molecule increases monotonically as the wave packet spreads.

\section{Entanglement of Coherent States and the Ground State}

The ground state and coherent states of the diatomic Hamiltonian are Gaussian states in either the molecular or the atomic basis, and as such their entanglement properties can be specified by the corresponding covariance matrix.  The symplectic eigenvalues of the partial transpose of a covariance matrix provide the separability criterion and can be used to calculate several measures of entanglement, such as logarithmic negativity~\cite{adesso_entanglement_2007,serafini_standard_2007}.  At the end of this section, we will apply these methods.  However, since our eventual goal is to evaluate (\ref{eq:pur}) for general, non-Gaussian states, it will be instructive to first consider the purity $P_{\alpha\beta}$ of molecular (two-mode) coherent states $\kt{\alpha,\beta}$.

In terms of the position coordinates $\Phi_{\alpha\beta}(r,x)=\bk{r,x}{\alpha,\beta}$, these states have Gaussian wave functions
\begin{equation}
\Phi_{\alpha\beta}(r,x)=\left(\frac{\gamma\Gamma}{\pi}\right)^{1/2}e^{-\frac{i}{2\hbar}(\alpha_r\alpha_q+\beta_x\beta_p)}e^{-\frac{\gamma^2}{2}(r-\alpha_r)^2-\frac{\Gamma^2}{2}(x-\beta_x)^2}e^{\frac{i}{\hbar}(r\alpha_q+x\beta_p)}
\end{equation}
where
\begin{eqnarray}\label{offset}
\alpha_r = \frac{1}{\sqrt{2}\gamma}(\alpha+\alpha^*),\ &&
\alpha_q = -\frac{i\hbar\gamma}{\sqrt{2}}(\alpha-\alpha^*),\nonumber\\
\beta_x =  \frac{1}{\sqrt{2}\Gamma}(\beta+\beta^*),\ &&
\mbox{and}\ \beta_p = -\frac{i\hbar\Gamma}{\sqrt{2}}(\beta-\beta^*).
\end{eqnarray}
Transforming to the particle coordinates via the symplectic transformation (\ref{eq:Observables Definition}), the new wave function $\tilde{\Phi}_{\alpha\beta}(x_1,x_2)=\Phi_{\alpha\beta}(x_1-x_2,\mu_1 x_1 + \mu_2 x_2)$ is still Gaussian.  The integral (\ref{eq:pur}) can be rewritten in the form
\begin{equation}\label{eq:cohint}
P_{\alpha\beta} = \frac{\gamma^2 \Gamma^2}{\pi^2} \int d^4{\bf z}e^{-{\bf z}^{\rm T}{\rm A}{\bf z}+{\bf B}^{\rm T}{\bf z}+C}
\end{equation}
with ${\bf z}=(x_1,x'_1,x_2,x'_2)^{\rm T}$, $d^4{\bf z}=dx_1dx'_1dx_2dx'_2$, $y=1/2(-\gamma^2 + \Gamma^2\mu_1\mu_2)$ and
\begin{eqnarray}
{\rm A} &=& \left(\begin{array}{cccc}
\Gamma^2 \mu_1^2 + \gamma^2 & 0 & y& y\\
0 & \Gamma^2 \mu_1^2 + \gamma^2  & y& y\\
 y& y& \Gamma^2 \mu_2^2 + \gamma^2 & 0 \\
 y& y&0& \Gamma^2 \mu_2^2 + \gamma^2 \end{array}\right)\label{AAA}\\
{\bf B} &=& 2\left(\begin{array}{c} \gamma^2 \alpha_r + \Gamma^2 \mu_1 \beta_x\\ \gamma^2 \alpha_r  + \Gamma^2 \mu_1 \beta_x\\ -\gamma^2 \alpha_r + \Gamma^2 \mu_2 \beta_x\\ -\gamma^2 \alpha_r  + \Gamma^2 \mu_2 \beta_x\end{array}\right)\\
C &= & -2(\gamma^2\alpha_r^2 + \Gamma^2\beta^2_x).
\end{eqnarray}
The integral (\ref{eq:cohint}) is standard when ${\rm A}$ is positive semidefinite:
\begin{equation}
\int d^n{\bf z}e^{-{\bf z}^{\rm T}{\rm A}{\bf z}+{\bf B}^{\rm T}{\bf z}+C} =
 \sqrt{\frac{\pi^n}{{\rm det}{\rm A}}}e^{\frac{1}{4}{\bf B}^{\rm T}{\rm A}^{-1}{\bf B} + C}.
\end{equation}
Therefore, making the necessary algebraic simplifications and noting ${\bf B}^{\rm T}{\rm A}^{-1}{\bf B}=-4C$, the purity of a coherent state is found to be
\begin{equation}\label{eq:coh}
P_{\alpha\beta} = \frac{\gamma\Gamma}{\sqrt{(\gamma^2 + \Gamma^2\mu_1^2)(\gamma^2 + \Gamma^2\mu_2^2)}}.
\end{equation}
This result shows that the entanglement for a two-mode coherent state in the center-of-mass/relative coordinates does not depend on the complex displacements $\alpha$ and $\beta$, as expected.  Since the displacement operators $D(\alpha,a)$ and $D(\beta,b)$ are separable in the atomic coordinates, then every coherent state must have the same entanglement as the ground state $\kt{0,0}$.  As a consequence, the purity of the ground state $P_{00}$ is also given by (\ref{eq:coh}).   

Figure 3 depicts the coherent state purity as functions of the kinematic parameter $\mu_1$ ($\mu_2=1-\mu_1$) and dynamical parameter $g=\omega/\Omega$:
\begin{equation}
P_{\alpha\beta}(g,\mu_1) =P_{00}(g,\mu_1)= \sqrt{ \frac{g}{(g\mu_1 + \mu_2)(g\mu_2 + \mu_1)}}.
\end{equation}
When the trap and molecular frequencies are the same ($g=1$), there is no atomic entanglement for any value of the mass ratios, as one might expect from the separability of the Hamiltonian.  Note that either transformation $g\rightarrow g^{-1}$ or $\mu_1\rightarrow \mu_2$ leaves $P_{00}$ unchanged.  As $g$ departs from one, the purity decreases and the entanglement grows without limit.  Maximum entanglement for a given $g$ occurs when the masses are equal $\mu_1=\mu_2=1/2$.  When the mass ratios of the two atoms are far off balance, the entanglement decreases.
\begin{figure}
\centering
\includegraphics{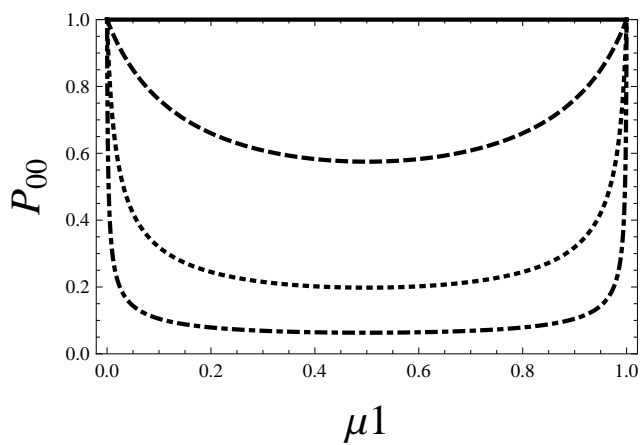}
\caption{The ground state/coherent state entanglement $P_{00}$ as a function of $\mu_1$ for four different values of $g=\omega/\Omega$: $g=1$ (solid), $g=10$ (dashed), $g=100$ (dotted), and $g=1000$ (dot-dashed).}
\label{fig3} 
\end{figure}

For the case of an unbound molecule, the purity of the time-dependent bound molecule $\tilde{\Phi}(t)_{0u}$ with center-of-mass momentum uncertainty $\hbar\Gamma$ at $t=0$ and relative vibrational state $m=0$ can be calculated the same way as the bound case, except the matrix ${\rm A}$ (\ref{AAA}) is now time dependent and takes the form
\begin{equation}\label{AAAt}
{\rm A(t)} = \left(\begin{array}{cccc}
\Gamma^2 \mu_1^2 + \gamma^2 & 0 & z^*& z\\
0 & \Gamma^2 \mu_1^2 + \gamma^2  &z& z^*\\
 z^*& z& \Gamma^2 \mu_2^2 + \gamma^2 & 0 \\
 z& z^*&0& \Gamma^2 \mu_2^2 + \gamma^2 \end{array}\right)
\end{equation}
where $z=-\gamma^2/2 + 1/2 e^{i\phi}\Gamma^2 \mu_1\mu_2$ and $\phi=\tan^{-1} \tau$.  Using this and following the same steps as above, we find that
\begin{equation}
P_{0u}(t)= \frac{\gamma\Gamma}{\sqrt{(\gamma^2 + \Gamma^2\mu_1^2)(\gamma^2 + \Gamma^2\mu_2^2)+\gamma^4 \tau^2}}.
\end{equation}
Note that $P_{0u}(t)$ has its maximum value at the moment of minimum uncertainty $t=M \tau/(\hbar \Gamma^2) =0$ and then decreases, meaning entanglement increases as time evolves.  An interesting relation to note is that when $\gamma=\Gamma\sqrt{\mu_1\mu_2}$, there is no entanglement between the atoms in the free molecule at time $t=0$. Figure 4 depicts the entanglement at $t=0$ using the  parameterization $c=\Gamma/\gamma$ instead of $g=\omega/\Omega$ to highlight these features.

\begin{figure}
\centering
\includegraphics{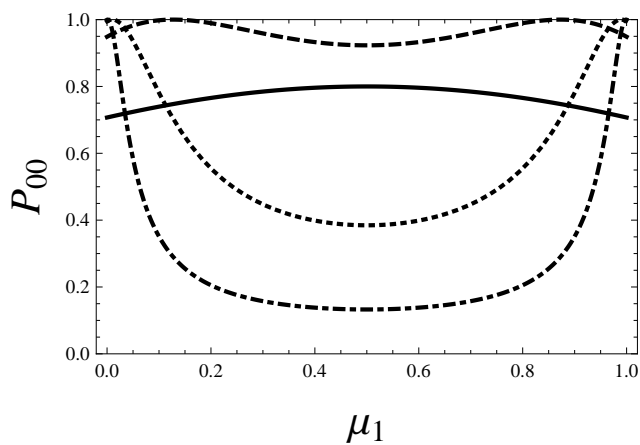}
\caption{The entanglement $P_{0u}(t)$ of $\tilde{\Phi}(t)_{0u}$ at time $t=0$ as a function of $\mu_1$ for four different values of $c=\gamma/\Gamma$: $c=1$ (solid), $c=3$ (dashed), $c=10$ (dotted), and $c=30$ (dot-dashed).}
\label{fig4}
\end{figure}

To conclude this section, we compare these results for Gaussian states to results from the covariance matrix approach.  Gaussian states are fully characterized by their first and second moments, but as we have shown, the first moments have no influence on entanglement properties since they can be removed by a local unitary transformation.  Defining the vector of operators
\begin{equation}
\hat{\bf R}=\left(\begin{array}{c}\sqrt{2}(
\hat{X}_1-\langle \hat{X}_1 \rangle) \\ \frac{\sqrt{2}}{\hbar}(\hat{P}_1-\langle \hat{P}_1 \rangle)\\ \sqrt{2}(\hat{X}_2-\langle \hat{X}_2 \rangle) \\ \frac{\sqrt{2}}{\hbar}(\hat{P}_2-\langle \hat{P}_2 \rangle)\end{array}\right),
\end{equation}
the elements of the covariance matrix can be calculated from $\mathrm{V}_{kl}=\langle \hat{R}_k \hat{R}_l + \hat{R}_l \hat{R}_k\rangle/2$.  For either the ground state or the coherent state of the bound molecule, the covariance matrix evaluates to
\begin{equation}\label{VVV}
\mathrm{V}=\left( \begin{array}{cccc} \frac{1}{\Gamma^2} + \frac{\mu_2^2}{\gamma^2} & 0 &  \frac{1}{\Gamma^2} - \frac{\mu_1\mu_2}{\gamma^2} & 0\\
0 & \gamma^2 + \Gamma^2 \mu_1^2 & 0 & -\gamma^2 + \Gamma^2 \mu_1 \mu_2\\
\frac{1}{\Gamma^2} - \frac{\mu_1\mu_2}{\gamma^2} & 0 & \frac{1}{\Gamma^2} + \frac{\mu_1^2}{\gamma^2} & 0\\
0 &  -\gamma^2 + \Gamma^2 \mu_1 \mu_2 & 0 & \gamma^2 + \Gamma^2 \mu_2^2 \end{array}\right).
\end{equation}
A symplectic transformation to a new covariance matrix $\mathrm{V}'= \mathrm{S}\mathrm{V}\mathrm{S}^\top$ with $\mathrm{S}\in\mathrm{Sp}(4,\mathbb{R})$ of the form $\mathrm{S}=\mathrm{S}_2 \oplus \mathrm{S}_2$ ($\mathrm{S}_2\in\mathrm{Sp}(2,\mathbb{R})$) is local with respect to the atomic observables and will not change the entanglement properties of the covariance matrix.  In particular, one can define a symplectic transformation that rescales the variables as
\begin{equation}
\mathrm{S}  = \left( \begin{array}{cc} \sqrt{\gamma\Gamma} s & 0\\
 0& 1/\sqrt{\gamma\Gamma} s \end{array}\right) \oplus \left( \begin{array}{cc} \sqrt{\gamma\Gamma}/s & 0\\
 0& s/\sqrt{\gamma\Gamma} \end{array}\right)
\end{equation}
with $s^4=(\gamma^2 + \Gamma^2\mu_1^2)/(\gamma^2 + \Gamma^2\mu_2^2)$.  This transformation brings the covariance matrix $\mathrm{V}'$
into the standard form for a two-mode squeezed state~\cite{serafini_standard_2007}
\begin{equation}
\mathrm{V}'=\left( \begin{array}{cccc} \cosh r & 0 & \sinh r & 0\\
0 & \cosh r & 0 & -\sinh r\\
\sinh r & 0 & \cosh r & 0\\
0 &  -\sinh r & 0 & \cosh r \end{array}\right).
\end{equation}
where the squeezing parameter $r$ is directly related to the purity $P_{00}$ by $\cosh r = (P_{00})^{-1}$.  For comparison with other results, note that for the covariance matrix $\mathrm{V}'$ one finds that the logarithmic negativity $\mathcal{E}_\mathcal{N}$  (a standard measure of entanglement for Gaussian states~\cite{adesso_entanglement_2007})  is exactly the squeezing parameter $\mathcal{E}_\mathcal{N} = r $.  A similar result (but with a more complicated,  time-dependent symplectic transformation $\mathrm{S}$) holds for the unbound state $\tilde{\Phi}(t)_{0u}$.

\section{Entanglement in Number States and Superpositions of Number States}

The purity for number states can be calculated using the connection between coherent states and number states
\begin{equation}\label{numtocoh}
\kt{n} = \left.\frac{1}{\sqrt{n!}}\frac{\partial^n}{\partial \alpha^n} e^{\frac{|\alpha|^2}{2}}\kt{\alpha}\right|_{\alpha=0}.
\end{equation}
By substituting the wave functions for coherent state in atomic coordinates into the expression (\ref{numtocoh}), we find
\begin{equation}\label{numtocoh2}
\tilde{\Phi}_{m,n}(x_1,x_2)= \left.\frac{1}{\sqrt{m!n!}}\frac{\partial^m}{\partial \alpha^m}\frac{\partial^n}{\partial \beta^n} e^{\frac{|\alpha|^2+|\beta|^2}{2}}\Phi_{\alpha\beta}(x_1-x_2,\mu_1 x_1 + \mu_2 x_2)\right|_{\alpha,\beta=0}\!\!\!\!\!\!\!\!.
\end{equation}

From here we can proceed in two ways.  We can substitute (\ref{numtocoh2}) directly into the purity expression (\ref{eq:pur}); that will be done below as part of the main text.  An alternate approach uses an expansion onto another double harmonic oscillator basis, one that is separable in atomic coordinates and denoted $\left|j,k\right\}$:
\begin{eqnarray}\label{dhb12}
\{ x_1,x_2 |j,k\}&=&\{ x_1|j\}\{ x_2 |k\}\nonumber\\
&=&\left(\frac{\gamma_1\gamma_2}{2^{j+k}\pi j!k!}\right)^{1/2} H_j(\gamma_1 x_1) H_k(\gamma_2 x_2)e^{-\gamma_1^2 x_1^2/2-\gamma_2^2 x_2^2/2}.
\end{eqnarray}
In this expression, $\gamma_1$ and $\gamma_2$ do not have a dynamical meaning based on the Hamiltonian like $\gamma$ and $\Gamma$, but instead are free parameters that should cancel out in the final expression for the purity.  Applying (\ref{numtocoh}) to the atomic basis vectors $\left|j,k\right\}$, we can find expressions for the coefficients $\langle n,m \left|j,k\right\}$ that transform between the molecular number basis to the (artificial) atomic number basis.  More details on this approach, which may be more useful for numerical simulations and for calculating the entropy of entanglement or other entanglement measures, are located in Appendix A.

Proceeding by direct substitution of (\ref{numtocoh2}) into the purity expression (\ref{eq:pur}) and performing the integral yields 
\begin{equation}\label{monstera}
P_{mn}=\left.\frac{\gamma^2 \Gamma^2}{(\pi m!n!)^2}\sqrt{\frac{\pi^4}{{\rm det}{\rm A}}}\left(\prod_{i=1}^4 \frac{\partial^m}{\partial \alpha_i^m}\frac{\partial^n}{\partial \beta_i^n}\right) e^{\frac{1}{4}{\bf B}^{\rm T}{\rm A}^{-1}{\bf B} + C}\right|_{\{\alpha_i,\beta_i\}=0}
\end{equation}
The real symmetric matrix ${\rm A}$ is the same as (\ref{AAA}) above, but now
\begin{equation}
{\bf B}=\sqrt{2}\left(\begin{array}{c}
\gamma (\alpha_1+\alpha_2) + \Gamma \mu_1 (\beta_1+\beta_2)\\
\gamma (\alpha_3+\alpha_4) + \Gamma \mu_1 (\beta_3+\beta_4)\\
-\gamma (\alpha_1+\alpha_4) + \Gamma \mu_2 (\beta_1+\beta_4)\\
-\gamma (\alpha_3+\alpha_2) + \Gamma \mu_2 (\beta_3+\beta_2)
\end{array}\right)
\end{equation}
and 
\begin{equation}
C= -1/2(\alpha_1^2 + \alpha_2^2 + \alpha_3^2 + \alpha_4^2+\beta_1^2 + \beta_2^2 + \beta_3^2 + \beta_4^2).
\end{equation}
Simplification leads to
\begin{equation}\label{monster}
P_{mn}=\left.\frac{P_{00}}{( m!n!)^2}\left(\prod_{i=1}^4 \frac{\partial^m}{\partial \alpha_i^m}\frac{\partial^n}{\partial \beta_i^n}\right) e^{{\bf z}^{\rm T}{\rm M}{\bf z}}\right|_{\{\alpha_i,\beta_i\}=0},
\end{equation}
where
\begin{equation}
{\bf z}^{\rm T}=(\alpha_1,\alpha_2,\alpha_3,\alpha_4,\beta_1,\beta_2,\beta_3,\beta_4)
\end{equation}
and ${\rm M}$ is an $8\times 8$ matrix (\ref{mmm}) described in Appendix B.

The expression (\ref{monster}) can be evaluated analytically for all values of $m$ and $n$, but it grows in complexity rapidly.  Explicit calculations reveal that all $P_{mn}$ have the form
\begin{equation}
\frac{P_{00}}{\left[(\gamma^2 + \Gamma^2\mu_1)(\gamma^2 + \Gamma^2\mu_2)\right]^{2m+2n}}\sum_{i=0}^{2m+2n}C^{(mn)}_i\gamma^{2i}\Gamma^{4m+4n-2i}
\end{equation}
where $C^{(mn)}_i$ are polynomials of $\mu_1$ and $\mu_2$ with rational coefficients that can be determined from ${\rm M}$.  As examples, one finds
\begin{eqnarray}
P_{01} &=& \frac{\gamma\Gamma}{4\left[(\gamma^2 +\mu_1^2\Gamma^2)(\gamma^2 +\mu_2^2\Gamma^2)\right]^{5/2}}\times\left(3\gamma^8+4\gamma^6\Gamma^2(\mu_1^2 +\mu_2^2)\right.\nonumber\\
& & +\left.2\gamma^4\Gamma^4(2\mu_1^4 + \mu_1^2\mu_2^2 + 2\mu_2^4) + 4\gamma^2\Gamma^6\mu_1^2\mu_2^2(\mu_1^2+\mu_2^2)+3\Gamma^8\mu_1^4\mu_2^4\right)
\end{eqnarray}
and
\begin{eqnarray}
P_{11} &=& \frac{\gamma\Gamma}{16\left[(\gamma^2 +\mu_1^2\Gamma^2)(\gamma^2 +\mu_2^2\Gamma^2)\right]^{9/2}} \left(9\gamma^{16}+16\gamma^{14}\Gamma^2(\mu_1^2 +\mu_2^2) \right.\nonumber\\
&&\left. + 12 \gamma^{12} \Gamma^4 (8\mu_1^4 -3 \mu_1^2\mu_2^2 + 8\mu_2^4)
+  240 \gamma^{10} \Gamma^6\mu_1^2\mu_2^2(\mu_1^2+\mu_2^2)  \right.\nonumber\\
 &&\left. + 2 \gamma^8 \Gamma^8 (8 \mu_1^8 -64 \mu_1^6\mu_2^2 + 459\mu_1^4\mu_2^4 -64 \mu_1^2\mu_2^6 + 8\mu_2^8)\right.\nonumber\\
&&\left. + 240 \gamma^6 \Gamma^{10}\mu_1^4\mu_2^4(\mu_1^2+\mu_2^2) 
+ 12 \gamma^4 \Gamma^{12} \mu_1^4\mu_2^4(8\mu_1^4 -3 \mu_1^2\mu_2^2 + 8\mu_2^4)\right.\nonumber\\
&&\left. +16\gamma^2\Gamma^{14}\mu_1^6\mu_2^6(\mu_1^2 +\mu_2^2)
+ 9\Gamma^{16}\mu_1^8\mu_2^8\right).
\end{eqnarray}
As for coherent states, one can rewrite any $P_{mn}$ so it depends only on the ratio of the frequencies $g$ (or equivalently momentum uncertainty ratio $c$) and the mass fraction $\mu_1$ (or $\mu_2$).  The lowest nine combinations of $m,n$ are depicted in Figure 5.
\begin{figure}
\centering
\includegraphics[width=\linewidth]{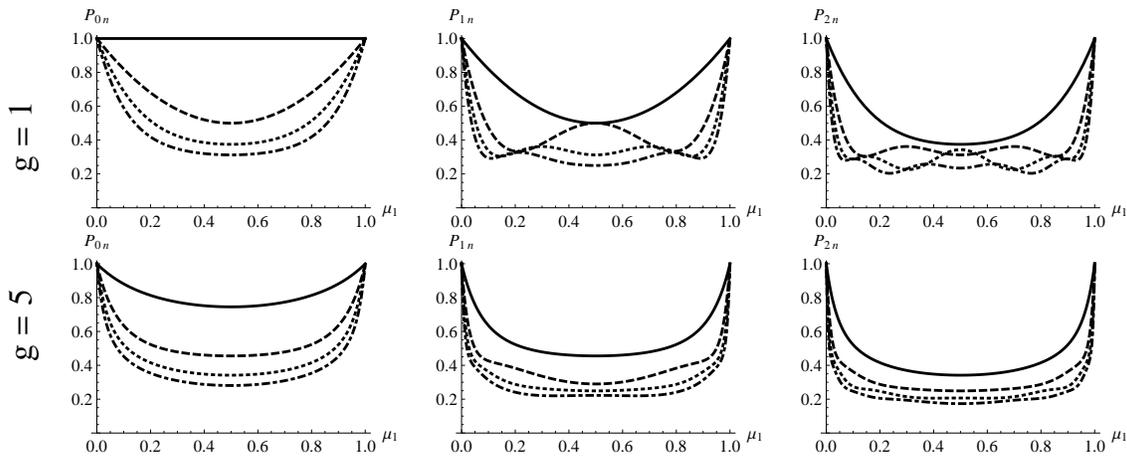}
\caption{Each column displays the graphs of $P_{mn}$ for $m=\{0,1,2\}$ (left, middle, right) and $n=\{0,1,2,3\}$ (solid, dashed, dotted, dot-dashed) for two values of $g$: $g=1$ (top) and $g=5$ (bottom).}
\label{fig5} 
\end{figure}

The following properties of the functions $P_{mn}$ can be inferred either by analytical means or by graphical inspection of the calculated purity for the lowest combinations of $m,n$:
\begin{itemize}

\item The purity functions are symmetric under exchange of trap and molecular quantum number, i.e. $P_{mn}=P_{nm}$.  Similar to the ground state purity $P_{00}$, the transformations  $\mu_1\rightarrow\mu_2$ or $g\rightarrow g^{-1}$ leave $P_{mn}$ invariant, as can be expected from the symmetries of the Hamiltonian.

\item For all $\{m,n\}$, the energy eigenstates have no entanglement in the limiting cases of extreme mass difference, either $m_2\gg m_1$ ($\mu_1\approx 0$) or $m_2\ll m_1$ ($\mu_1\approx 1$).  In this unphysical limit, the length scale $\gamma^{-1} = \sqrt{\hbar/\mu \omega}$ diverges and the energy eigenstates $\tilde{\Phi}_{m,n}(x_1,x_2)$ become unnormalizable.  

\item For all finite mass ratios, one finds $P_{mn}<1$ and therefore the energy eigenstates are entangled.  The only exception is the special case $g=1$ and then only the ground state is separable.  Inspecting the Hamiltonian in atomic coordinates (\ref{eq:atoHam}) one can see that  operator becomes separable whenever $g = 1$.  However, this does not imply that the energy eigenfunctions $\tilde{\Phi}_{m,n}(x_1,x_2)$ become separable in that limit.  As a side note, this does mean that for uncoupled oscillators there exists an alternate energy eigenstate basis $\kt{m,n}$ constructed of entangled states that coincides with the atomic oscillator basis $|j,k\}$ only on the ground state.

\item When $g=1$, $P_{mn}$ is a polynomial of $\mu_1$ of order $2(m+n)$.  For example, $P_{00}=1$, $P_{10}=P_{01}=1-2\mu_1 + 2\mu_1^2$, and $P_{11}= 1 - 8 \mu_1 + 32  \mu_1^2 - 48  \mu_1^3 + 24  \mu_1^4$.

\item Based on graphical analysis, for a fixed value of $g$ it appears $P_{m+1,n+1}<P_{m,n}$ for all $\mu_1$, but no other inequalities appear universal.  For example, there are some regions of $\{g,\mu_1\}$-parameter space where $P_{12}>P_{11}$ and where $P_{13}>P_{12}$.  As $m$ and $n$ increase the functions become more oscillatory and more tightly spaced, making it unlikely to hypothesize any other bounds based on graphical methods alone.

\item For a fixed value of $\mu_1$, the entanglement generally increases as $g$ increases.  As $g\rightarrow \infty$, the purity takes its global minimum at $\mu_1 = 1/2$, although for general $g$, $P_{mn}$ need not have a minimum at $\mu_1=1/2$, and it may in fact have a local maximum.  For fixed $\mu_1$, the entanglement increase  is not monotonic in $g$, but shows local maxima and minima of decreasing prominence as $g$ increases.  

\end{itemize}
As before, one can use the same method, only replacing ${\rm A}$ (\ref{AAA}) with ${\rm A(t)}$ (\ref{AAAt}), to calculate $\tilde{\Phi}(t)_{mu}$ for unbound diatomic molecules with center-of-mass momentum uncertainty $\hbar\Gamma/\sqrt{2}$ and vibrational state $m$.

Finally, in principle one can  calculate the entanglement of general states
\begin{equation}
\kt{\Phi} = \sum_{m,n=0}^\infty c_{m,n} \kt{m,n}
\end{equation}
as
\begin{equation}
P(\Phi)=\sum_{\{m_i,n_i\}_{i\in\{1,2,3,4\}}=0}^{\infty} c_{m_1,n_1} c^*_{m_2,n_2}c_{m_3,n_3} c^*_{m_4,n_4}P(\{m_i,n_i\})
\end{equation}
where
\begin{equation}
P(\{m_i,n_i\}) = \frac{P_{00}}{\prod_{i=1}^4 m_i!n_i!}\left(\prod_{\{\alpha,\beta\}_i=1}^4 \frac{\partial^{m_i}}{\partial \alpha_i^{m_i}}\frac{\partial^{n_i}}{\partial \beta_i^{n_i}}\right) \left.e^{{\bf z}^{\rm T}{\rm M}{\bf z}}\right|_{\{\alpha_i,\beta_i\}=0}
\end{equation}
In contrast to the purities of number states, $P_{mn}$, the functions $P(\{m_i,n_i\})$ are not necessarily positive or symmetric around $\mu_1=1/2$.  Nor do they limit to unity when $\mu_1\rightarrow 0$ or 1, although numeric analysis suggests they are bounded functions.  Many symmetry relations between permutations of indices can be derived. For example, because of the integration range, only even kernels contribute, meaning $P(\{m_i,n_i\})=0$ unless $\sum_{i=1}^4 (m_i+n_i)$ is even.

Figure 6 gives a flavor for the entangle\-ment properties of mole\-cular state super\-positions.  The purity of the state
\begin{equation}\label{sup}
\kt{\Phi} = \cos\theta \kt{0,1} + \sin\theta\kt{1,0}
\end{equation}
is depicted as a function of $\mu_1$ for several values of $g$ and $\theta$.

\begin{figure}
\centering
\includegraphics{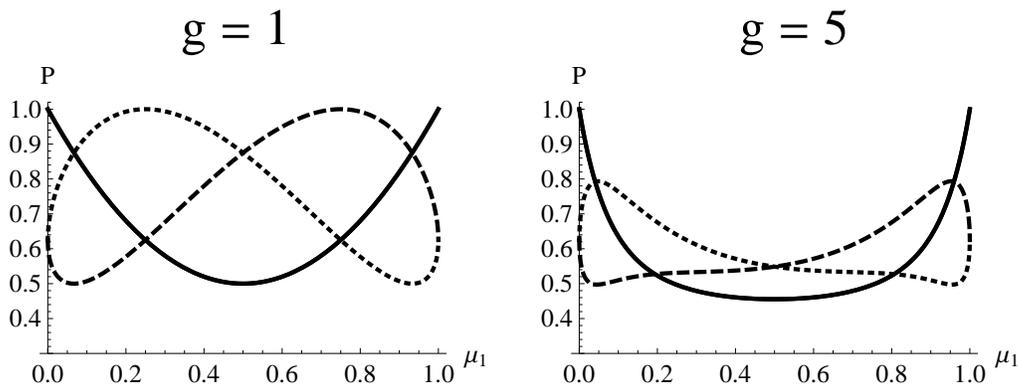}
\caption{This figure depicts the purity of the reduced density matrix of (\ref{sup}) for two values of $g$ and three values of $\theta$: $\theta=n\pi/2$ (solid), $\theta=n\pi/2+\pi/6$ (dashed), and $\theta=n\pi/2+\pi/3$ (dashed) for $n$ any integer.  Note that for $g=1$ and $\theta\neq n\pi/2$, there are certain mass ratios that have no interatomic entanglement.}
\label{fig6} 
\end{figure}

\section{Conclusions: Correlation, Covariance, and the Classical-Quantum Correspondence}

When the trap frequency and molecular frequency are the same ($g=1$), there is no entanglement between the atoms when the center-of-mass and relative oscillators are in coherent states, and in particular the ground state. For all other coherent states and number states with finite mass ratios, there is entanglement between the atoms, which generally increases with $g$ and with the index of the number states. Certain linear combinations of number states, including those entangled with respect to the molecular observables,  can be disentangled with respect to atomic observables, but again apparently only when $g=1$.  These mathematical results, based on the definition of separability with respect to a given tensor product structure, can be mathematically proven.

In interacting systems, correlations are expected, and one perspective is that this kind of entanglement is just an artifact of studying the molecule in the `wrong' basis, i.e.\ the atomic basis.  Since the Hamiltonian is separable with respect to the center-of-mass/relative basis, in some sense the dynamics `chooses' the molecular observables over the particle observables.  However, if the atoms have internal structure, one can imagine at least in principle, that the atomic observables could be experimentally accessed (or ``chosen'') independently  of the molecular observables and attempts could be made to measure correlations between the atoms.
For example, one could attempt to measure the covariance between $\hat{X}_1$ and $\hat{X}_2$:
\begin{equation}
\sigma_{x_1 x_2}={\rm V}_{13}/2 = \langle \hat{X}_1\hat{X}_2\rangle - \langle \hat{X}_1\rangle\langle \hat{X}_2\rangle.
\end{equation}
For molecular coherent states, one finds
\begin{equation} 
\sigma_{x_1 x_2} = \frac{1}{2\Gamma^2} - \frac{\mu_1\mu_1}{2 \gamma^2} = \frac{1}{2\Gamma^2}(1-g^{-1}).
\end{equation}
This covariance quantifies the correlations in uncertainty that the atomic positions inherit from the intrinsic minimum uncertainty of the molecular oscillators.  Low frequency traps with high frequency molecules imply the largest covariance (and therefore easiest to measure).  When $g=1$, the covariance disappears for coherent states, as does the entanglement, which can be seen from (\ref{VVV}).  The same holds for the other elements of the covariance matrix.  However, although there is entanglement, one can show that these correlations are not intrinsically quantum.  We could imagine a classical analogue: two masses on a spring.  The masses could be at rest, but with statistical uncertainty in their positions, so that the probability distribution as a function of center of mass coordinates $\{x,p\}$ and relative coordinates $\{r,q\}$ is
\begin{equation}
\rho(x,p,r,q)= \frac{1}{\hbar^2\pi^2} e^{-\Gamma^2 (x-x_0)^2}e^{-\frac{1}{\hbar^2 \gamma^2} (p-p_0)^2}e^{-\gamma^2 (r-r_0)^2}e^{-\frac{1}{\hbar^2 \gamma^2} (q-q_0)^2}.
\end{equation}
This supposition gives the same covariance matrix ${\rm V}$ as the quantum coherent state.  The only difference between the quantum and classical covariance is that in the quantum case $\gamma$ and $\Gamma$ are of dynamical origin, whereas they are of purely statistical origin in the classical case.  Another way of saying this is that the transformation from molecular observables to atomic observables maps a positive, Gaussian Wigner function into another positive Gaussian Wigner function, and one can show that some classical model can reproduce correlations in any positive Wigner function.

In contrast, in the number state $\kt{m,n}$ the position covariance $\sigma_{x_1 x_2}$ is
\begin{eqnarray} 
\sigma_{x_1 x_2} &=& \frac{1}{2\Gamma^2}(2n+1) - \frac{\mu_1\mu_1}{2 \gamma^2}(2m+1)\nonumber\\
&=& \frac{1}{2\Gamma^2}((2n+1)-g^{-1}(2m+1)).
\end{eqnarray}
This has very different properties from the coherent state covariance, and from the entanglement of such states.  When $n=m$ and for $g=1$ the covariance in number states disappears, although the entanglement is non-vanishing for all $m$ and $n$ except $m=n=0$.  The covariance is greatest in magnitude for $g\gg 1$ and large $n$ or $g\ll 1$ and large $m$, a relationship between $g$ and number that does not exist for the entanglement.  For number states there is no correspondence to a classical model with statistical uncertainty.  Additionally, we note that the Wigner function for a number state has negative regions, often considered a signal of `quantumness', whether expressed in atomic or molecular variables, but entanglement only in atomic variables.

  As a final comment, we note that the expressions for purity do not depend on $\hbar$ or the ratio of the quantum scale to the classical scale in any way. Although measurements of the postion and momentum uncertainties will involve a scale set by $\hbar$, in the purity expression only the ratios appear and so the overall scale cancels out.  To see whether the entanglement correlations are truly ``quantum'' one could also imagine constructing dichotomous observables on the $(x_1,x_2)$-space such that Bell-type inequalities can be formulated.  Such observables can be constructed in several ways, for example based on the displaced parity operator~\cite{banaszek_nonlocality_1998} or using pseudospin operators~\cite{chen_maximal_2002}, and this investigation will be pursued in future work.

\ack

Both authors would like to thank the Research Corporation for its support and Joshua Lansky and Michael Keynes for enlightening conversations.  W.F.F also acknowledges the Barry M.~Goldwater Scholarship and Excellence in Education Program, and N.L.H. expresses gratitude to the Deutscher Akademischer Austausch Dienst for supporting his visit to the Institute of Quantum Physics at the University of Ulm, hosted by Wolfgang Schleich and Matthias Freyberger.  Finally, N.L.H. thanks his hosts in Ulm and Lev Plimak for valuable discussions.

\appendix

\section{An Alternate Approach}

If the goal is to calculate the entanglement of a state with respect to the atomic tensor product structure $\HS=\HS_1\otimes\HS_2$, then any basis that is separable with respect to this structure can be used for taking the partial trace.  In the main body of this paper, the purity of the reduced density matrices was calculated using the continuous-variable atomic coordinate basis $|x_1,x_2\}$.  In this appendix, we instead use a double harmonic oscillator basis $\left|j,k\right\}$.  These states are realized by separable wave functions (\ref{dhb12}) characterized by positive real parameters $\gamma_1$ and $\gamma_2$ that can be freely chosen for convenience.  There can be advantages of using such a discrete basis to generate approximate expressions for the purity even when an exact analytic expression can also be derived.

Using the $\left|j,k\right\}$ basis, the reduced density matrix for atom 1 can be written
\begin{equation}
\hat{\rho}_1=\sum_{j,j',k=0}^\infty \left\{j,k\right|\Phi\rangle \langle \Phi\left|j',k\right\}\left|j\right\}\left\{j'\right|
\end{equation}
and the purity of the reduced density matrix is 
\begin{equation}\label{purdis}
P(\Phi)=\sum_{j,j',k,k'=0}^\infty \left\{j,k\right|\Phi\rangle \langle \Phi\left|j',k\right\}\left\{j',k'\right|\Phi\rangle \langle \Phi\left|j,k'\right\}.
\end{equation}

We will focus on calculating the entanglement for number states $\kt{\Phi}=\kt{m,n}$, and the matrix elements transforming between the molecular oscillators and the atomic oscillators $\left\{j,k\right|m,n\rangle$ are the central objects of concern.  Similar to the procedure in the main text, an expression for this matrix element will be derived by taking the derivatives of the matrix element between coherent states $\left\{\tau_1,\tau_2\right|\alpha,\beta\rangle$:
\begin{eqnarray}
\left\{j,k\right|m,n\rangle &&= \frac{1}{\sqrt{j!k!m!n!}}\label{jkmn}\\
&&\times \frac{\partial^{j+k+m+n}}{\partial \tau_1^j\partial\tau_2^k\partial\alpha^m\partial\beta^n}\left. e^{\frac{1}{2}(|\tau_1|^2+|\tau_2|^2+|\alpha|^2+|\beta|^2)}\left\{\tau_1,\tau_2\right|\alpha,\beta\rangle\right|_{\tau_1,\tau_2,\alpha,\beta=0}\!\!\!\!\!\!\!\!\!\!\!\!\!\!\!\!\!\!\!\!\!\!.\nonumber
\end{eqnarray}
The molecular coherent state $\kt{\alpha,\beta}$ is defined above (\ref{molcoh}) and the atomic coherent states are defined as
\begin{equation}\label{atocoh}
\left|\tau_1,\tau_2\right\} = e^{-|\tau_1|^2/2 -|\tau_2|^2/2}\sum_{j,k=0}^{\infty}\frac{\tau_1^j\tau_2^k}{\sqrt{j! k!}} \left|j,k\right\}.
\end{equation}
The associated wave function in particle coordinates is
\begin{eqnarray}
\langle x_1, x_2 \left|\tau_1,\tau_2\right\} &=& \left(\frac{\gamma_1\gamma_2}{\pi}\right)^{1/2} e^{-\frac{i}{2\hbar}(\tau_{1x}\tau_{1p}+\tau_{2x}\tau_{2p})} \nonumber\\
&&\times e^{-\frac{\gamma_1^2}{2} (x_1 -\tau_{1x})^2 -\frac{\gamma_2^2}{2} (x_2 -\tau_{2x})^2 }e^{\frac{i}{\hbar}(\tau_{1p}x_1+\tau_{2p}x_2)},
\end{eqnarray}
with analogs definitions for $\tau_{1x}$, $\tau_{1p}$, etc., to (\ref{offset}).
One can then evaluate $\left\{\tau_1,\tau_2\right|\alpha,\beta\rangle$ by performing the integral
\begin{equation}
\left\{\tau_1,\tau_2\right|\alpha,\beta\rangle = \int{dx_1 dx_2 \left\{\tau_1,\tau_2\kt{x_1,x_2}\langle{x_1,x_2}\right|\alpha,\beta\rangle}.
\end{equation}
This is yet another Gaussian integral, but this time only in two variables.  Completing the integration and substituting into (\ref{jkmn}), the coefficient can be written
\begin{eqnarray}
\left\{j,k\right|m,n\rangle& =& \left(\frac{4\gamma_1\gamma_2\gamma\Gamma}{\sqrt{j!k!m!n!}Z}\right)^{1/2}\nonumber\\
&&\times \left.\frac{\partial^{j+k+m+n}}{\partial \tau_1^j\partial\tau_2^k\partial\alpha^m\partial\beta^n} e^{-1/2(\tau_1^2+\tau_2^2+\alpha^2+\beta^2)}e^{F/Z}\right|_{\tau_1,\tau_2,\alpha,\beta=0},\label{jkmn_ev}
\end{eqnarray}
where
\begin{eqnarray}
F &=& (\gamma^2 +\mu_2^2\Gamma^2 +\gamma_2^2)(\gamma \alpha + \mu_1\Gamma\beta +\gamma_1\tau_1)^2\nonumber\\
&& + 2(\gamma^2-\mu_1\mu_2\Gamma^2)(\gamma \alpha + \mu_1\Gamma\beta +\gamma_1\tau_1)(-\gamma \alpha + \mu_2\Gamma\beta +\gamma_2\tau_2)\nonumber\\
&& +(\gamma^2 +\mu_1^2\Gamma^2 +\gamma^2_1)(-\gamma \alpha + \mu_2\Gamma\beta +\gamma_2\tau_2)^2\nonumber\\
Z &=& \gamma^2\Gamma^2 + \mu_2^2\gamma_1^2\Gamma^2+\mu_1^2\gamma_2^2\Gamma^2 + \gamma_1^2\gamma_2^2 + \gamma^2(\gamma_1^2 + \gamma_2^2).
\end{eqnarray}

The expression (\ref{jkmn_ev}) depends in a complicated fashion on the non-physically meaningful parameters $\gamma_1$ and $\gamma_2$, but surprisingly, when this coefficient is substituted into the summation in (\ref{purdis}), this dependence must cancel.  To examine how this sum converges to the exact result (\ref{monster}), Figure 7 depicts the convergence of $P_{01}$ for four cases of $(g,\mu_1)$ and for four several different values of $(\gamma_1,\gamma_2)$.  We do not answer the potentially interesting question of how to choose $\gamma_1$ and $\gamma_2$ for optimal convergence, although some features make intuitive sense.  For example, for smaller $g$, smaller choices for $(\gamma_1,\gamma_2)$ converge faster.  Also, for the cases where $\mu_1=1/10$, choices with $\gamma_1<\gamma_2$ converge faster than those with $\gamma_1>\gamma_2$.

\begin{figure}
\centering
\includegraphics{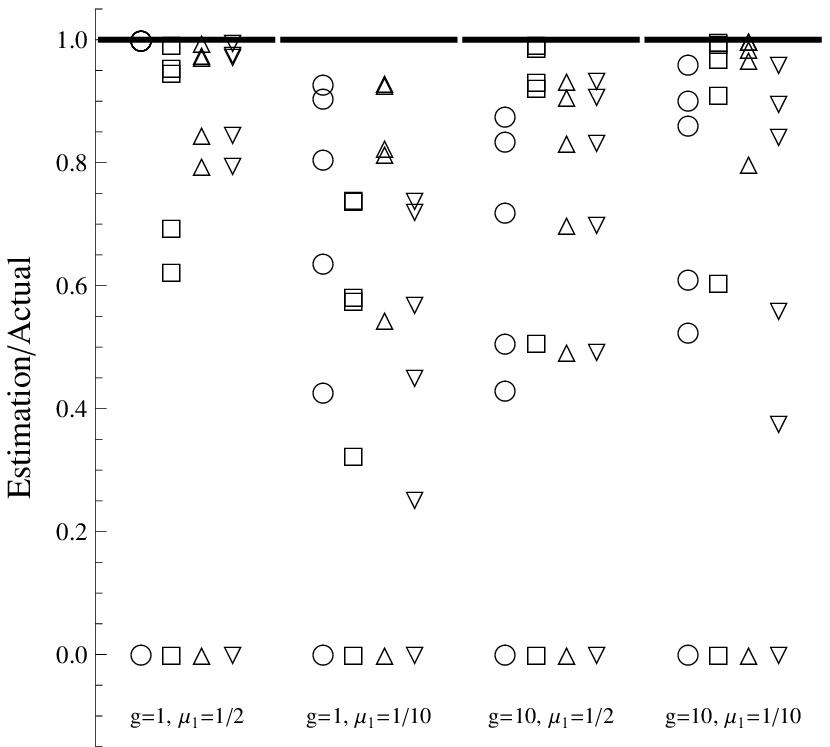}
\caption{This figure depicts the convergence of the approximation scheme for $P_{01}$ for four different values of $(g,\mu_1)$ and four different choices of the parameters $(\gamma_1,\gamma_2)$: $(\gamma_1,\gamma_2)=(1/\sqrt{2},1/\sqrt{2})$ (circle), $(1,1)$ (square), $(1/\sqrt{2},1)$ (up-triangle), and $(1,1/\sqrt{2})$ (down-triangle).  For each combination of $(g,\mu_1)$ and $(\gamma_1,\gamma_2)$, the accuracy  of six successive approximations to $P_{01}$ are plotted. Each approximation corresponds to taking more and more terms in the sum in (\ref{purdis}) from $j_{\rm{max}}=k_{\rm{max}}=0$ to $5$.  If not all six shapes are apparent, successive approximation give results indistinguishable on this scale.  For the first case, $(g,\mu_1)=(1,1/2)$, the first choice $(\gamma_1,\gamma_2)=(1/\sqrt{2},1/\sqrt{2})$ gives the exact result at the second approximation $j_{\rm{max}}=k_{\rm{max}}=1$.}
\label{fig7} 
\end{figure}

The advantage of using this method is that the reduced density matrix $\hat{\rho}_1$ can be approximated to arbitrary accuracy by a finite-dimensional matrix.  This matrix, for example, could be diagonalized and used to calculate the entropy of entanglement.  Also, the coefficients $\left\{j,k\right|m,n\rangle$ are necessary if one is to use the pseudo-spin operators~\cite{chen_maximal_2002} to construct Bell-type inequalities for this system.  More generally, even though one can find exact analytic continuous-variable expressions for the purity of the reduced matrix elements, numerical schemes require discretization, which in certain cases has been shown to mask the presence of entanglement in continuous variable systems~\cite{harshman_entanglement_2008-1}.

\section{The matrix $\rm{M}$}

The entanglement of an number state can be calculated exactly using the expression
\begin{equation}\label{monsterapp}
P_{mn}=\left.\frac{P_{00}}{( m!n!)^2}\left(\prod_{i=1}^4 \frac{\partial^m}{\partial \alpha_i^m}\frac{\partial^n}{\partial \beta_i^n}\right) e^{{\bf z}^{\rm T}{\rm M}{\bf z}}\right|_{\{\alpha_i,\beta_i\}=0},
\end{equation}
where  
\begin{equation}
{\bf z}^{\rm T}=(\alpha_1,\alpha_2,\alpha_3,\alpha_4,\beta_1,\beta_2,\beta_3,\beta_4).
\end{equation}
The matrix $\rm{M}$ is $8\times 8$ and can be written as
\begin{equation}\label{mmm}
\rm{M} = \left(\begin{array}{cccccccc}
u & v & -u & w & s & -t & -s & t\\
v & u & w & -u & -t & s & t & -s\\
-u & w & u & v & -s & t & s & -t\\
w & -u & v & u & t & -s & -t & s\\
s & -t & -s & t & -u & w & u & v\\
-t & s & t & -s & w & -u & v & u\\
-s & t & s & -t & u & v & -u & w\\
t & -s & -t & s & v & u & w & -u
\end{array}\right)
\end{equation}
where
\begin{eqnarray*}
u&=&\gamma^4 - \Gamma^4\mu_1^2\mu_2^2/D\\
v&=& \gamma^4 +2\gamma^2\Gamma^2\mu_1^2 + \Gamma^4\mu_1^2\mu_2^2/D\\
w&=&\gamma^4 +2\gamma^2\Gamma^2\mu_2^2 + \Gamma^4\mu_1^2\mu_2^2/D\\
s&=& \gamma\Gamma(\gamma^2  -\Gamma^2 \mu_1\mu_2)(\mu_1 -\mu_2)/D\\
t&=& \gamma\Gamma(\gamma^2  +\Gamma^2 \mu_1\mu_2)(\mu_1 +\mu_2)/D=\gamma\Gamma(\gamma^2  +\Gamma^2 \mu_1\mu_2)/D\\
D&= &4(\gamma^2 + \Gamma^2\mu_1)(\gamma^2 + \Gamma^2\mu_2)
\end{eqnarray*}
It is useful to note that $\det{M}=1/256$.


\section*{References}


\begin{thebibliography}{99}

\bibitem{james_quantum_1998}
James D 1998 {\em Applied Physics B: Lasers and Optics\/} {\bf 66} 181
  
\bibitem{zhu_trapped_2006}
Zhu S, Monroe C and Duan L 2006 {\em Physical Review Letters\/} {\bf 97} 050505
 
\bibitem{audenaert_entanglement_2002}
Audenaert K, Eisert J, Plenio M~B and Werner R~F 2002 {\em Physical Review A\/}
  {\bf 66} 042327


\bibitem{amico_entanglement_2008}
Amico L, Fazio R, Osterloh A and Vedral V 2008 {\em Reviews of Modern
  Physics\/} {\bf 80} 517
 

\bibitem{adesso_entanglement_2007}
Adesso G and Illuminati F 2007 {\em Journal of Physics A: Mathematical and
  Theoretical\/} {\bf 40} 7821

\bibitem{serafini_standard_2007}
Serafini A and Adesso G 2007 {\em Journal of Physics A: Mathematical and
  Theoretical\/} {\bf 40} 8041
 

\bibitem{fan_eigenvectors_1994}
Fan H and Klauder J~R 1994 {\em Physical Review A\/} {\bf 49} 704

\bibitem{simon_peres-horodecki_2000}
Simon R 2000 {\em Physical Review Letters\/} {\bf 84} 2726 

\bibitem{botero_modewise_2003}
Botero A and Reznik B 2003 {\em Physical Review A\/} {\bf 67} 052311
  

\bibitem{marian_gaussian_2008}
Marian P and Marian T~A 2008 {\em The European Physical Journal - Special
  Topics\/} {\bf 160} 281
 
\bibitem{pirandola_correlation_2009}
Pirandola S, Serafini A and Lloyd S 2009 {\em Physical Review A\/} {\bf 79} 052327

\bibitem{zanardi_quantum_2004}
Zanardi P, Lidar D~A and Lloyd S 2004 {\em Physical Review Letters\/} {\bf 92}
  060402 

\bibitem{torre_entanglement_2010}
de~la Torre A~C, Goyeneche D and Leitao L 2010 {\em European Journal of
  Physics\/} {\bf 31} 325

\bibitem{hongyi_relationship_2003}
Fan H and Yue F 2003 {\em Journal of Physics A: Mathematical and General\/}
  {\bf 36} 5319

\bibitem{fan_hong-yi_common_1995}
Fan H and Xiong Y 1995 {\em Physical Review A\/} {\bf 51} 3343 

\bibitem{fan_solvingtwo-body_1996}
Fan H and Chen B 1996 {\em Physical Review A\/} {\bf 53} 2948 

\bibitem{schmuser_entanglement_2006}
Schmuser F and Janzing D 2006 {\em Physical Review A \/} {\bf 73} 052313

\bibitem{harshman_entanglement_2008-1}
Harshman N~L and Singh P 2008 {\em Journal of Physics A: Mathematical and
  Theoretical\/} {\bf 41} 155304 

\bibitem{fedorov_packet_2004}
Fedorov M~V, Efremov M~A, Kazakov A~E, Chan K~W, Law C~K and Eberly J~H 2004
  {\em Physical Review A\/} {\bf 69} 052117 

\bibitem{fedorov_spontaneous_2005}
Fedorov M~V, Efremov M~A, Kazakov A~E, Chan K~W, Law C~K and Eberly J~H 2005
  {\em Physical Review A\/} {\bf 72} 032110 

\bibitem{fedorov_short-pulse_2006}
Fedorov M~V, Efremov M~A, Volkov P~A and Eberly J~H 2006 {\em Journal of
  Physics B: Atomic, Molecular and Optical Physics\/} {\bf 39} S467 

\bibitem{jost_entangled_2009}
Jost J~D, Home J~P, Amini J~M, Hanneke D, Ozeri R, Langer C, Bollinger J~J,
  Leibfried D and Wineland D~J 2009 {\em Nature\/} {\bf 459} 683

\bibitem{serafini_manipulatingquantum_2009}
Serafini A, Retzker A and Plenio M~B 2009 {\em New Journal of Physics\/} {\bf
  11} 023007 

\bibitem{banaszek_nonlocality_1998}
Banaszek K and Wodkiewicz K 1998 {\em Physical Review A\/} {\bf 58} 4345

\bibitem{chen_maximal_2002}
Chen Z, Pan J, Hou G and Zhang Y 2002 {\em Physical Review Letters\/} {\bf 88}
  040406 

\bibitem{gneiting_entanglingfree_2009}
Gneiting C and Hornberger K 2009 
  \texttt{http://arxiv.org/abs/0905.1279}

\bibitem{moshinsky_good_1968}
Moshinsky M 1968 {\em American Journal of Physics\/} {\bf 36} 52

\bibitem{oneill_wave_2003}
{O'Neill} D~P and Gill P~M~W 2003 {\em Physical Review A\/} {\bf 68} 022505
 

\bibitem{march_proposed_2008}
March N~H, Cabo A, Claro F and Angilella G~G~N 2008 {\em Physical Review A
  \/} {\bf 77} 042504

\bibitem{neal_density_1998}
Neal H~L 1998 {\em American Journal of Physics\/} {\bf 66} 512

\bibitem{ragot_exact_2006}
Ragot S 2006 {\em The Journal of Chemical Physics\/} {\bf 125} 014106

\bibitem{jens_peder_dahl_moshinsky_2009}
Dahl J~P 2009 {\em Canadian Journal of Chemistry\/} {\bf 87} 784

\bibitem{amovilli_quantum_2004}
Amovilli C and March N~H 2004 {\em Physical Review A\/} {\bf 69} 054302

\bibitem{march_kinetic_2006}
March N~H, Negro J and Nieto L~M 2006 {\em Journal of Physics A: Mathematical
  and General\/} {\bf 39} 3741

\bibitem{yaez_quantum_2010}
Ya\~nez R, Plastino A and Dehesa J 2010 {\em The European Physical Journal D\/}
  {\bf 56} 141

\bibitem{coe_entanglement_2008}
Coe J~P, Sudbery A and {D'Amico} I 2008 {\em Physical Review B \/} {\bf 77} 205122

\bibitem{pipek_measures_2009}
Pipek J and Nagy I 2009 {\em Physical Review A \/} {\bf 79} 052501

\bibitem{schleich_quantum_2001}
Schleich W~P 2001 {\em Quantum Optics in Phase Space\/} 1st ed ({Wiley-VCH})
  

\end{thebibliography}
\end{document}